\documentclass[useAMS,usenatbib]{mn2e}
\usepackage{amssymb}
\usepackage{amsmath}
\usepackage{graphicx}
\def\hi{H\,{\sc i}}
 
\def\deg{$^{\circ}$}
\def\kms{km~s$^{-1}$}
\def\msun{M$_{\odot}$}
\long\def\symbolfootnote[#1]#2{\begingroup%
\def\thefootnote{\fnsymbol{footnote}}\footnote[#1]{#2}\endgroup} 

\def\aj{AJ}%
\def\apj{ApJ}%
\def\apjl{ApJ}%
\def\apss{Ap\&SS}%
\def\aap{A\&A}%
\def\aapr{A\&A~Rev.}%
\def\mnras{MNRAS}%

\title[A search for non-circular flows in the extended \hi\ disc of NGC~2915]{A search for non-circular flows in the extended \hi\ disc of NGC~2915}
\author[Elson et al.]{E. C. Elson$^{1}$\thanks{Current address: South African Astronomical Observatory, PO~Box~9, Observatory 7935, South Africa}\thanks{E-mail:
elson.e.c@gmail.com (ECE), edeblok@ast.uct.ac.za (WJGdeB), kraan@ast.uct.ac.za (RCK-K)}, W. J. G. de Blok$^{1}$ and R. C. Kraan-Korteweg$^{1}$\\
$^{1}$Astrophysics, Cosmology and Gravity Centre (ACGC), Department of Astronomy, University of Cape Town, Private Bag X3,\\
Rondebosch 7701, South Africa\\}
\begin{document}

\date{Accepted 2010 September 8.}

\pagerange{\pageref{firstpage}--\pageref{lastpage}} \pubyear{2002}

\maketitle

\label{firstpage}

\begin{abstract}
NGC~2915 is a nearby blue compact dwarf with a differentially rotating \hi\ disc extending out to $\sim 5$ $R$-band $R_{25}$ radii.  This disc serves as an ideal tracer of the system's gravitational potential in regions of the galaxy that are dominated by dark matter.  We use new \hi\ synthesis observations of NGC~2915, obtained with the Australia Telescope Compact Array, to search for non-circular flows within the outer \hi\ disc.  Two independent methods are used, and the results of each interpreted in the context of relevant axisymmetric and non-axisymmetric perturbations of the potential.  We find evidence for: (1) elliptical streaming associated with the spiral structure of the \hi\ disc and the central bar-like feature in the mass distribution, (2) a spherical dark matter halo, and (3) an axisymmetric radial outflow of $\sim 5$-17~\kms\ ($\sim6$-20~percent of the circular speed).  A possible bar-like perturbation of the potential hinders attempts to unambiguously detect kinematic signatures of radial flows in the \hi\ velocity field.  The radial outflows are inconsistent with the plausible disc formation scenario in which gas from the surrounding inter-galactic medium is deposited on the outer  \hi\ disc and then transported towards the centre of the galaxy.  They are, however, consistent with the possibility of some material being re-distributed towards the outer disc in order to conserve angular momentum as material flows inwards along a bar.  
\end{abstract}


\section{Introduction}\label{intro}
Observed velocity fields of late-type galaxies are often treated as being consistent with a purely circular flow pattern.  However, the intrinsic, and therefore also the observed, properties of a galaxy's kinematics will be affected by the presence of non-circular velocity components within the orbits of observable test particles.  These non-circular motions are caused in various ways, and can be either systematic or random in nature.  On small length scales ($\lesssim$ 0.5 kpc) for example, stellar feedback can significantly disrupt gas dynamics, setting up expanding shells within a galaxy's inter-stellar medium (ISM) \citep[e.g.][]{IC2574}.  On larger scales, stellar feedback can drive galactic outflows in systems of sufficiently low dynamical mass \citep{meurer_1705_1992,martin_IZW18_1996,maclow_1999,van_eymeren}.  Globally, a non-axisymmetric gravitational potential will force orbits to be intrinsically elliptical rather than circular \citep{binney_warps_1978,bosma_thesis}. 

Given an observed velocity field, it is the task of the investigator to determine what information the observed flow pattern contains about the mass distribution and other properties of the system.  In the case where the orbits of tracer particles are significantly affected by non-circular motions, a careful characterisation of the non-circular signatures present within the velocity field is required.  Several authors \citep[e.g.][]{franx_et_al_1994,schoenmakers_1997,canzian_1997,spekkens_sellwood_2007} have developed methods for investigating and interpreting complex, non-circular velocity fields of galaxies in the context of perturbations of the potential.  These methods allow further insight into the fundamental nature of the processes responsible for generating the non-circular components, as well as the non-circular components themselves.  As an example, \citet{wong} used the CO and \hi\ velocity fields of seven nearby spirals to search for evidence of radial gas flows.  They found deviations from pure circular rotation at the level of 20-60~\kms\ and attributed them to the effects of bar streaming, inflows and warps.  More recently, \citet{trachternach_THINGS} quantified the non-circular motions in a sample of 19 galaxies from the The \hi\ Nearby Galaxy Survey \citep[THINGS,][]{THINGS_walter}.  They found a median absolute amplitude of the non-circular motions, averaged over their entire sample, of 6.7~\kms.  From the measured non-circular velocity components they determined the mean elongation of the gravitational potential to be $0.017\pm 0.020$, consistent with a round potential.

\begin{figure}
	\begin{centering}
	\includegraphics[angle=0,width=1\columnwidth]{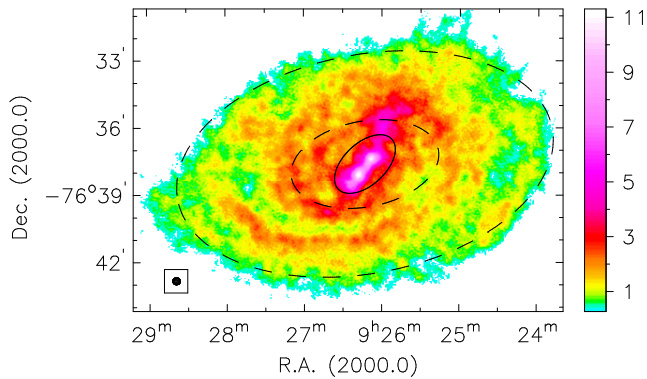}
	\caption{\hi\ surface density map of NGC~2915 from E2010a.  The logarithmic intensity scale is in units of \msun~pc$^{-2}$.  The solid ellipse with semi-major axis length 98~arcsec represents the $R$-band $R_{25}$ radius.  The dashed ellipses have semi-major axis lengths of 200~arcsec and 510~arcsec, and delimit the portion of the \hi\ disc for which an harmonic decomposition of the line-of-sight velocities is carried out in this paper.}
	\label{mom0}
	\end{centering}
\end{figure}

In this work we search for non-circular flows in the extended \hi\ disc of the nearby \citep[$4.1\pm0.3$~Mpc, ][]{meurer2003} galaxy, NGC~2915.  Optically, the galaxy is classified as a blue compact dwarf.  Centred on the optical disc \citep[$R_{25}\sim 98$~arcsec in the $R$-band,][]{meurer1} is a differentially rotating \hi\ disc with well-defined spiral structure, extending further than $\sim 500$~arcsec.  We use new \hi\ synthesis observations \citep[][hereafter E2010a]{elson_2010a_temp} to study the \hi\ kinematics of the galaxy.  The \hi\ total intensity map of NGC~2915 is shown in Fig~\ref{mom0}.  In this paper we are primarily concerned with the search for and the physical interpretation of non-circular velocity components within the \hi\ disc.  Two independent methods are used to quantify the non-circular flows.  The results are interpreted in the context of various axisymmetric and non-axisymmetric perturbations of the potential.  The first method uses the approach developed by \citet{spekkens_sellwood_2007} for fitting a general axisymmetric or non-axisymmetric model to the velocity field of a galaxy.  They develop the model specifically for the case of bi-symmetric distortions (bar- or oval-like) and for purely axisymmetric radial flows.  The second technique is that of \citet{franx_et_al_1994} and \citet[][]{schoenmakers_1997}, used to interpret the harmonic decomposition results of an observed velocity field in the context of time-varying perturbations to an axisymmetric potential.

The outline of this paper is as follows: the \hi\ data are introduced in Section~\ref{HI_data}.  The observed \hi\ velocity field of NGC~2915, along with a pure circular rotation model approximation of it, is shown in Section~\ref{resid_vel_field}.  We proceed to decompose and model the observed line-of-sight velocities in Section~\ref{method} using each of the two above-mentioned independent methods.  The various results are then interpreted, compared and discussed.  Section~\ref{epotential} focuses on the possible tri-axial nature of NGC~2915's dark matter halo.  Finally, our summary and conclusions are presented in Section~\ref{conclusions}.

\section{\hi\ data}\label{HI_data}
In this work we utilise the \hi-line data set of NGC~2915 presented in E2010a.  The \hi\ observations were carried out between 2006 October 23 and 2007 June 2 (project number C1629) with six different Australia Telescope Compact Array antenna configurations using all six antennas.  E2010a describe the various telescope setups of the observations as well as the reduction procedures used to arrive at the calibrated, continuum-subtracted, deconvolved \hi\ data cubes.  This work is based on their naturally-weighted \hi\ data cube which has spatial and spectral resolutions of 17.0~arcsec~$\times$~18.2~arcsec and 3.2~\kms\ respectively\symbolfootnote[1]{At the distance of NGC~2915 ($\sim 4.1$~Mpc), 1~arcsec corresponds to a linear distance of 19.8~pc.}.  We use the \hi\ velocity field which was extracted from the naturally-weighted \hi\ data cube by fitting a third-order Gauss-Hermite ($h_3$-GH) polynomial to each \hi\ line profile.  The reader is referred to \citet{franx_et_al_1994,noordermeer2007,THINGS_deblok} for various discussions of the merits of such a means of velocity field construction.  In the context of this work it suffices to say that the $h_3$-GH parameterisation of a line profile is less susceptible to the effects of profile asymmetries on circular velocity estimates made for a rotating disk.  
\section{Evidence of non-circular flows}\label{resid_vel_field}
\begin{figure}
	\begin{centering}
	\includegraphics[angle=0,width=1\columnwidth]{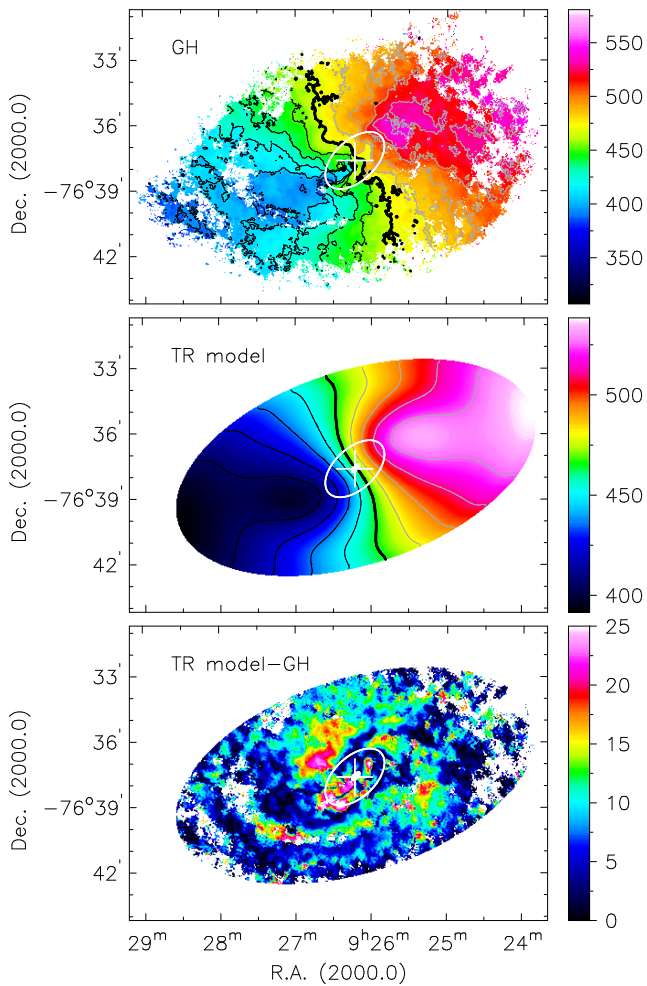}
	\caption{Top panel: $h_3$-GH velocity field from E2010a; Middle panel: model circular velocity field based on the tilted ring model from E2010a.  Bottom panel: absolute residual velocity field obtained by subtracting the model field  from the $h_3$-GH velocity field.  In all panels, contours are separated by 15~\kms, with the thick contour marking the systemic velocity at 465~\kms.  Grey and black contours represent receding and approaching halves of the galaxy respectively.  The intensity scale of the line-of-sight velocities is in units of \kms.  All panels show the same area.  The white ellipse represents the $R$-band $R_{25}$ radius of 98~arcsec of the stellar disc.  The white cross represents the photometric centre determined my E2010a.}
\label{velocityfields}	
	\end{centering}
\end{figure}

The $h_3$-GH \hi\ velocity field of NGC~2915 is shown in the top panel of Fig.~\ref{velocityfields}.  The general pattern of the iso-velocity contours is that of a disc that is rotating linearly at inner radii and differentially at outer radii.  The sharp kinks and wiggles along the iso-velocity contours, particularly prevalent at inner radii, are indicative of non-circular velocity components within the \hi.  E2010a fitted two different tilted ring models to this \hi\ velocity field.  The only significant difference between their two models was the best-fitting inclination profile.  Their so-called model CI had a constant inclination of $\sim55$\deg\ for all rings, while their model SI exhibited a sharp increase of $\sim 20$\deg\ when moving from the outer to the inner disc.  The models differed only slightly in terms of their circular velocity profiles (see E2010a for the full details and results of the modeling procedure).  

The middle panel of Fig.~\ref{velocityfields} shows a purely circular model velocity field generated using the best-fitting parameters of tilted ring model~CI from E2010a.  This model cannot adequately account for the observed velocities.  As a measure of the typical departures from pure circular motion at various positions within the galaxy, a residual velocity field (shown in the bottom panel of Fig.~\ref{velocityfields}) is produced by determining the absolute difference between the observed and model velocity fields.  From this residual map it is clear that the vast majority of the large residuals lie close to the centre of NGC~2915 and along the \hi\ spiral arms.  The mean absolute residual is $\sim 7.8$~\kms, with 69~percent of the area of the residual velocity field having a value less than 10~\kms\ (the approximate velocity dispersion of the outer \hi\ disc).  This suggests that a significant fraction of the \hi\ is regularly rotating in circular orbits about the kinematic centre. However, almost 9~percent of the residuals are larger than 15~\kms, some being as high as 20-25~\kms, indicative of significant non-circular \hi\ velocity components.



\section{Quantifying the non-circular flows}\label{method}
\subsection{Method 1: {\sc velfit}}
\subsubsection{Introduction}
The first method used to quantify the non-circular flows within the \hi\ disc of NGC~2915 is the procedure initially introduced by \citet{spekkens_sellwood_2007}.  Given certain assumptions regarding the nature of the non-circular flows, their fitting routine, {\sc velfit}, is able to fit to an observed velocity field a kinematic model describing the non-circular velocities.  The line-of-sight velocity, $V_{los}$, of a test particle orbiting the kinematic centre of a galaxy within the plane of the disc is
\begin{equation}
V_{los}=V_{sys}+\sin i(V_t\cos\theta+V_r\sin\theta),
\end{equation}
where $V_{sys}$ is the systemic velocity of the galaxy, $V_t$ and $V_r$ are the tangential and radial velocity components respectively, and $\theta$ is the angle in the disc plane relative to the projected major axis.  As explained by \citet{spekkens_sellwood_2007}, the above expression can be re-written in terms of a Fourier series:
\begin{align}
V_{los}&=V_{sys}+\sin i[V_t\cos\theta+\sum_{m=1}^{\infty}V_{m,t}\cos\theta\cos(m\theta+\theta_{m,t})\nonumber \\
&+V_r\sin\theta+\sum_{m=1}^{\infty}V_{m,r}\sin\theta\cos(m\theta+\theta_{m,r})],
\label{fourier_vel}
\end{align}
where $V_{m,t}$ and $V_{m,r}$ represent the magnitudes of the non-circular velocity components, and $\theta_{m,t}$ and $\theta_{m,r}$ their phases relative to some axis.  

{\sc velfit} makes the following assumptions in order to model $V_{m,t}$ and $V_{m,r}$ for a two-dimensional velocity field:
\begin{enumerate}
\item The non-circular motions are caused by a bar- or oval-like distortion to an otherwise axisymmetric potential.  Alternatively, the velocity field is treated as containing a radial flow component.  The harmonic order of the potential perturbation \emph{in the plane of the disc} is represented by $m$.\\
\item The $m=2$ terms dominate the non-circular velocity components for a strong bi-symmetric potential.  {\sc velfit} therefore only determines the $V_{2,t}$ and $V_{2,r}$ terms in Eqn.~\ref{fourier_vel}.  {\sc velfit} assumes that the bi-symmetric distortion has a fixed position angle, thereby yielding the fitted model insensitive to spiral structure.  Alternatively, {\sc velfit} can ignore all harmonic terms  with $m>0$, thereby fitting a purely axisymmetric radial flow model to the velocity field.\\
\item The disc under consideration is not warped.
\end{enumerate}
Assuming the $m=2$ term to dominate the non-circular velocity components, {\sc velfit} fits the following model velocity field to the observed velocity field:
\begin{align}
V_{model}=V_{sys}+\sin i[&V_t\cos\theta-V_{2,t}\cos(2\theta_b)\cos\theta \nonumber \\
-&V_{2,r}\sin(2\theta_b)\sin\theta],
\label{bisymm}
\end{align}
where $\theta_b$ is the major axis of the bi-symmetric distortion (e.g. a bar) relative to the projected major axis.  The routine determines radial profiles for $V_t$, $V_{2,t}$ and $V_{2,r}$ that allow for the best fit of the model velocity field to the observed velocity field (see Section~3.4 of \citet{spekkens_sellwood_2007} for a detailed discussion of the method by which {\sc velfit} determines the radial profiles).  The above-mentioned bi-symmetric model is also described by various disc parameters, the values of which {\sc velfit} estimates when fitting the model to the observed velocity field.  These additional parameters are: (1) $x_c$ and $y_c$, the coordinates of the kinematic centre of the disc; (2) $V_{sys}$; (3) $\epsilon_d$, the ellipticity of the disc; (4) $\phi'_d$, the major axis position angle of the disc in the sky plane; and (5) $\phi'_b$, the bar position angle in the sky plane.  Assuming all harmonic terms with $m>0$ to be negligible, {\sc velfit} can also fit a purely axisymmetric radial flow model to the velocity field:
\begin{equation}
V_{model}=V_{sys}+\sin i(V_t\cos\theta+V_r\sin\theta).
\label{radial}
\end{equation}

 \citet{spekkens_sellwood_2007}, as well as \citet{sellwood_zanmar_sanchez_2010}, highlight several features of the {\sc velfit} routine which they argue makes it better than other methods of estimating non-circular motions in rotating discs.  {\sc velfit} fits the entire velocity field at once, thereby allowing the effects of a mild bi-symmetric distortion to the potential to be detected in the flow pattern.  Weak kinematic signatures, they say, can be missed when fitting models to only a small portion of the velocity field.  They suggest further that considering only a bi-symmetric model allows for a less ambiguous identification of $m=2$ components within the observed velocity field.  Most importantly, they highlight the fact that their fitting routine does not rely on the assumption that the non-circular velocity components are small relative to the mean circular motion.  

\subsubsection{Fitted models}
{\sc velfit} was used to fit bi-symmetric and radial flow model velocity fields (Eqns.~\ref{bisymm} and \ref{radial} respectively) to the $h_3$-GH \hi\ velocity field of NGC~2915.  Models were applied to the 17~arcsec-resolution velocity field (Fig.~\ref{high_res_velfit_models}), as well as a 34~arcsec-resolution velocity field (Fig.~\ref{low_res_velfit_models}).  For all models, the parameters $x_c$, $y_c$, $V_{sys}$, $\epsilon_d$, $\phi'_d$ and $\phi'_b$ were allowed to vary freely.  {\sc velfit} allows the user to specify a galactocentric radius within which no pixels of the velocity field are fitted.  In this way a central region of the velocity field can be excluded from being modelled.  We, however, always applied our models from the very centres of the velocity fields.  Fitted pixels were sampled in ring widths of 17~arcsec.

Stable fits to the high-resolution \hi\ velocity field could not be obtained when modeling the entire disc.  The resulting $V_t$, $V_r$, $V_{2,t}$ and $V_{2,r}$ profiles exhibited large, non-systematic variations at radii $R\gtrsim 300$~arcsec.  The high-resolution velocity field was therefore only fitted out to $R=300$~arcsec.  \citet{sellwood_zanmar_sanchez_2010} describe in their Appendix~A an optional penalty that can be applied within {\sc velfit} to slightly smooth the radial variations of the fitted $V_t$, $V_r$, $V_{2,t}$ and $V_{2,r}$ profiles, thereby allowing for more stable modeling results.   They state 
\begin{equation}
\lambda={0.002\times V_{typ}^2\over \left({R_{max}\over \Delta R}\right)^4}
\end{equation}
as an appropriate softening parameter, where $V_{typ}$ is an estimate of the typical orbital velocity in the disc, $\Delta R$ is the ring spacing and $R_{max}$ is the maximum galactocentric radius of the fitted model.  Using $V_{typ}=85$~\kms, $\Delta R=17$~arcsec and $R_{max}=300$~arcsec, softening parameters of $\lambda_1=\lambda_2=1.5\times10^{-4}$ were used to slightly smooth the radial variations of the $V_t$, $V_r$, $V_{2,t}$ and $V_{2,r}$ profiles of the high-resolution \hi\ velocity field\symbolfootnote[8]{The magnitude of the penalty can be set independently for the axisymmetric and non-axisymmetric terms, we used the same penalty for both.}.  \citet{sellwood_zanmar_sanchez_2010} point out that the penalising effects of the softening parameters are insensitive to variations of a factor of a few about their values.  

Models were fitted to the entire low-resolution \hi\ velocity field, with no smoothing parameters applied.  For all models, the differences between the observed and modelled line-of-sight velocities were weighted by the inverse of the typical velocity dispersion of the outer \hi\ disc (10~\kms; E2010a).  The position angle in the sky plane of the bi-symmetric distortion (i.e. the \hi\ bar, clearly visible in the \hi\ total intensity map) was not allowed to freely vary in the bi-symmetric flow models, and was fixed to $\phi'_b=322$\deg.  This value corresponds to the dynamical estimate of the position angle for the central \hi\ disc produced by E2010a who fitted tilted ring models to the high-resolution \hi\ velocity field.  Our best-fitting bi-symmetric flow models were found to be virtually identical for variations of a few degrees of $\phi'_b$.  Non-parametric estimates of the true uncertainties in the estimated parameters were generated by applying a bootstrap method.  The method uses the spread of estimated parameter values from repeated fits to the observed \hi\ velocity field to yield the non-parametric estimates.  A full description of this method of error estimation in the context of {\sc velfit} is provided in Appendix~B of \citet{sellwood_zanmar_sanchez_2010}.

\subsubsection{Results}\label{velfit_results}

\begin{figure*}
	\begin{centering}
	\includegraphics[angle=0,width=2\columnwidth]{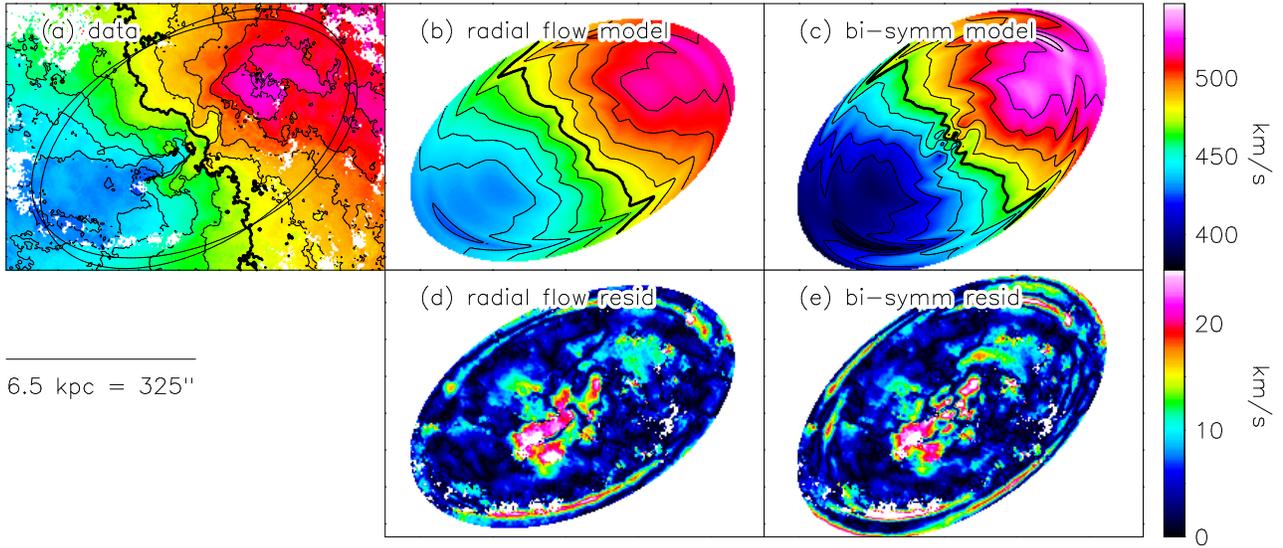}
	\caption{Kinematic models of the high-resolution NGC~2915 \hi\ velocity field.  Panels a, b, and c show the observed velocity field, the best-fitting axisymmetric radial flow model and the best-fitting bi-symmetric flow model respectively.  The ellipses in the top left panel delimit each of the model fields.  The velocity fields are all plotted using the same intensity scale in units of \kms.  Iso-velocity contours are spaced by 15~\kms, with the thick contour corresponding to 465~\kms.  Panels d and e show the respective residual maps.  Both maps are plotted using the same intensity scale in units of \kms.  Kinematic models were only fitted to radii $R<300$~arcsec, all maps therefore represent only this inner portion of the galaxy.  The portion of the observed velocity field shown in the top left panel is delimited by a black rectangle in Fig.~\ref{low_res_velfit_models}a.}
	\label{high_res_velfit_models}
	\end{centering}
\end{figure*}

\begin{figure*}
	\begin{centering}
	\includegraphics[angle=0,width=2\columnwidth]{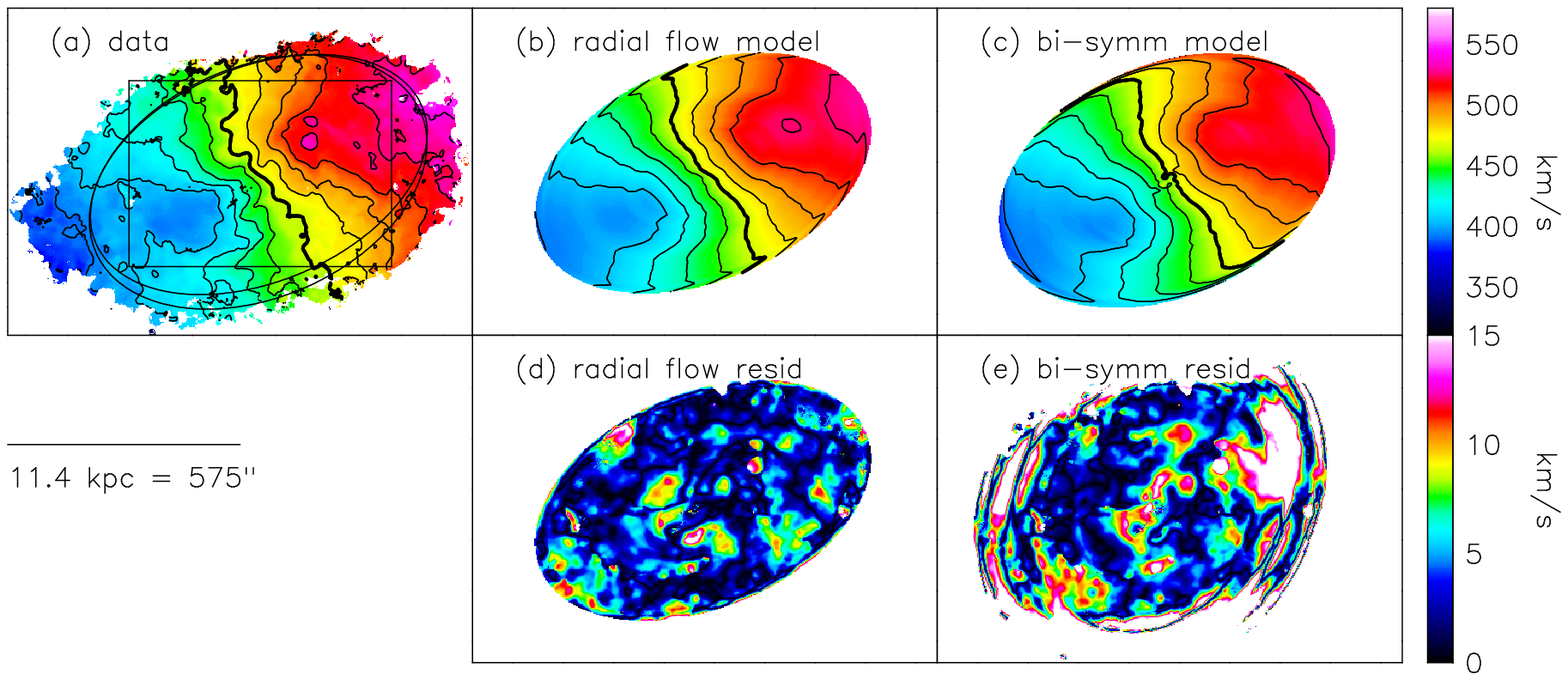}
	\caption{Kinematic models of the low-resolution NGC~2915 \hi\ velocity field.  Panels a, b, and c show the observed velocity field, the best-fitting axisymmetric radial flow model and the best-fitting bi-symmetric flow model respectively.  The ellipses in the top left panel delimit each of the model fields.  The velocity fields are all plotted using the same intensity scale in units of \kms.  Iso-velocity contours are spaced by 15~\kms, with the thick contour corresponding to 465~\kms.  Panels d and e show the respective residual maps.  Both maps are plotted using the same intensity scale in units of \kms.  Note the difference in scale between these maps and the maps presented in Fig.~\ref{high_res_velfit_models} above.  The black rectangle shown in panel a delimits the portion of the high-resolution \hi\ velocity field shown Fig.~\ref{high_res_velfit_models}a.}
	\label{low_res_velfit_models}
	\end{centering}
\end{figure*}

\begin{figure*}
	\begin{centering}
	\includegraphics[angle=0,width=2\columnwidth]{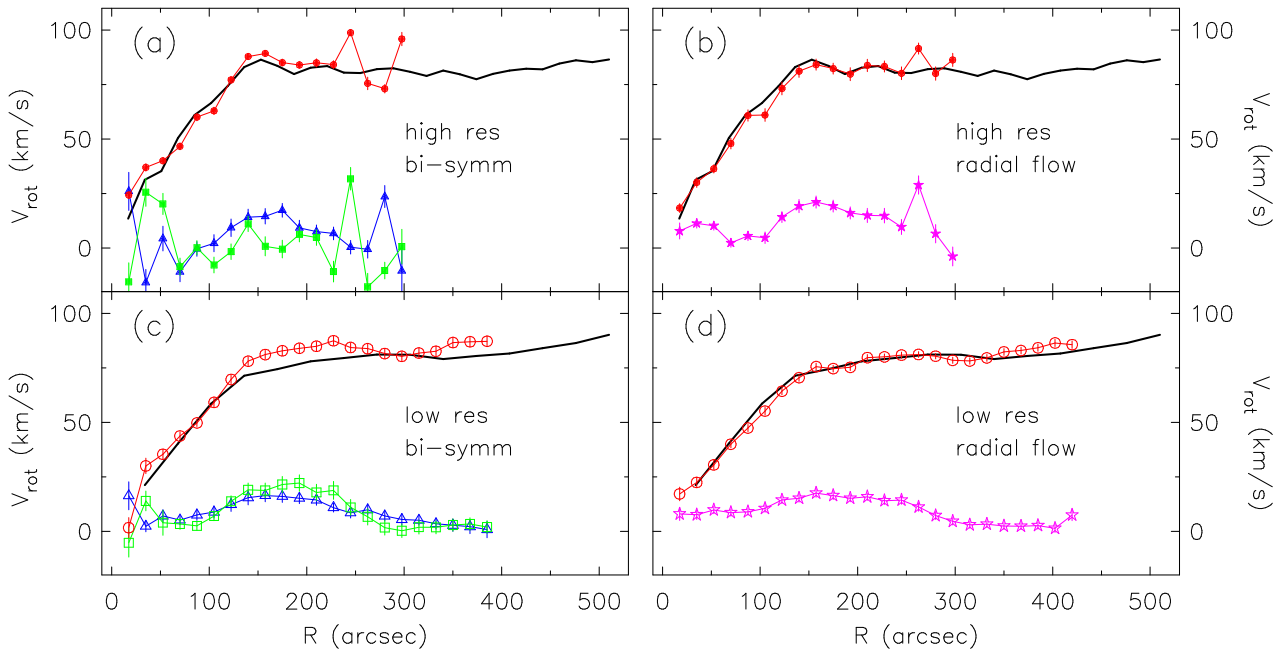}
	\caption{Fitted velocity components for NGC~2915 \hi\ velocity field.  In each panel the solid black curve represents the circular velocity profile of a tilted ring model fitted to the \hi\ velocity field.  The left-most panels show the $V_t$, $V_{2,t}$, $V_{2,r}$ components (red circles, green squares, and blue triangles respectively) of the best-fitting bi-symmetric flow models for the high- and low-resolution \hi\ velocity fields (top and bottom panels, filled and open symbols respectively).  The right-most panels show the $V_t$ and $V_r$ components (red circles and stars respectively) of the best-fitting axisymmetric radial flow models for the high- and low-resolution \hi\ velocity fields (top and bottom panels, filled and open symbols respectively).  In all panels, the error bars represent non-parametric estimates of the true uncertainties in the parameters, generated by applying a bootstrap method.}
	\label{velfit_profiles}
	\end{centering}
\end{figure*}

The final bi-symmetric and radial flow models of the high- and low-resolution observed \hi\ velocity fields are shown in Figs.~\ref{high_res_velfit_models} and \ref{low_res_velfit_models} respectively.  Also included in the figures are the corresponding absolute residual velocity fields, obtained by subtracting the models from the observed fields.  The best-fitting $x_c$, $y_c$, $V_{sys}$, $i$, $\phi'_d$ and $\phi'_b$ parameters for each of the final models are presented in Table~\ref{parameters_table}.  The $x_c$ and $y_c$ values represent the respective absolute $x$ and $y$ offsets of the fitted kinematic centres from the photometric centre of $\alpha_{2000}$~=~09$^\mathrm{h}$~26$^\mathrm{m}$~12.6$^\mathrm{s}$, $\delta_{2000}$~=~$-76$\deg~37$'$~37.8$''$, determined by E2010a.  The $V_t$, $V_{2,t}$, $V_{2,r}$ and $V_r$ radial profiles for the bi-symmetric and radial flow models are shown for the high- and low-resolution \hi\ velocity fields in Fig.~\ref{velfit_profiles}.  For comparison and for reference between plots, circular velocity profiles from tilted ring models fitted to each velocity field are presented as solid black curves in the various panels.  The circular velocity profile used for the high-resolution velocity field is from the so-called tilted ring model~CI presented in E2010a.  The best-fitting geometric and orientation parameters of this model were assumed for a tilted-ring model fitted to the low-resolution \hi\ velocity field, the circular velocity profile for which is shown in the lower panels of Fig.~\ref{velfit_profiles}.  

\begin{table*}
\begin{center}
\caption{Summary of best-fitting model parameters for NGC~2915.$^{a}$}
\label{parameters_table}
\begin{tabular}{cccccccccc}
\hline
\\
1&2&3&4&5&6&7&8&9&10\\ 
Model		&	$x_c$			&	$y_c$			&	$V_{sys}$		&	$i$	&	$\phi'_d$	&	$\phi'_b$	&	$\chi^2_{red}$	&	$\nu$		&	$<|\Delta V|>$\\
			&	(arcsec)			&	(arcsec)			&	(\kms)		&	(deg)	&	(deg)		&	(deg)		&				&			&	(\kms)\\
\\
\hline
\\
High res.	&&&&&&&&&\\ 
\\
Bi-symmetric	&	$1.1\pm 3.6$		&	$4.5\pm1.5$		&	$462.4\pm1.2$	&	$53.6\pm1.1$ 	&	$304.4\pm 2.2$	&	322.0		&	1.80	&	514	&	6.5\\
Radial		&	$1.9\pm 7.1$		&	$4.1\pm3.7$		&	$462.2\pm2.1$	&	$53.5\pm3.1$ 	&	$295.6\pm 3.3$	&	...		&	1.94	&	533	&	6.1\\
\hline
Low res.\\
\\
Bi-symmetric	&	$-5.2\pm 10.3$	&	$-20.5\pm5.5$		&	$458.5\pm1.7$	&	$49.1\pm3.0$ 	&	$293.3\pm 3.5$	&	322.0		&	0.71	&	1177	&	8.5\\
Radial		&	$3.7\pm 5.5$	&	$2.0\pm4.3$		&	$460.3\pm0.8$	&	$52.7\pm2.0$ 	&	$292.6\pm 2.2$	&	...		&	0.67	&	1202	&	3.7\\

\hline
 \end{tabular}
\end{center}
\begin{flushleft}
$^a$Column~1: type of model velocity field; Column~2/3: absolute $x$/$y$ offset of kinematic centre from the photometric centre; Column~4: systemic velocity; Column~5: disc inclination; Column~6/7: position angle of disc/bi-symmetric distortion; Column~8: minimum reduced $\chi^2$ statistic; Column~9: number of degrees of freedom when fitting; Column~10: average absolute residual.\\
Model parameters best-fitted to the high/low-resolution \hi\ velocity fields are listed in the top/bottom sections of the table.  Note that $\phi_b'$ is held fixed when fitting the bi-symmetric models, the only parameter with this distinction.
\end{flushleft}
\end{table*}

The overall kinematics of the models are consistent with those of the observations.  The iso-velocity contours of the bi-symmetric flow models of both the high- and low-resolution \hi\ velocity fields are sharper and more twisted at innermost radii than those of the corresponding radial flow models.  Despite being kinematically simpler than the bi-symmetric flow models, it is the radial flow models that best match the observations.  This is particularly evident in the case of the low-resolution \hi\ velocity field for which the radial flow model neatly reproduces the large-scale kinematic features.  The residual patterns for the high-resolution \hi\ velocity fields are similar for both models.  Residuals of $\sim 20$~\kms\ are observed at inner radii, and residuals $\lesssim 10$~\kms\ throughout the rest of the disc.  These smaller residuals are clumped together on roughly 80-100~arcsec angular scales ($\sim 1.6$-2.0~kpc).  Quite noticeably, the radial flow model of the low-resolution \hi\ velocity field allows for small residuals at inner radii, as well as  small residuals that are more uniformly distributed throughout the entire disc.  This is partly due the the lower angular resolution of the data, yet also indicates the accuracy of the fitted model.  Indeed, the average absolute residual corresponding to this model is only 3.7~\kms.  This model is also able to reproduce some of the closed iso-velocity contours seen in the north-west portion of the \hi\ disc.

The best-fitting disc parameters for both the high- and low-resolution \hi\ velocity fields are very similar to one another.  Furthermore, the fitted disc parameters are similar to those obtained by E2010A who fitted a tilted ring model to the high-resolution \hi\ velocity field.  The best-fitting inclinations of $\sim 53$\deg\ closely match their almost constant inclination of $\sim 55$\deg, while the systemic velocity estimates of $\sim460-462$~\kms\ agree well with their preferred 465~\kms.  The fitted kinematic centres for the high-resolution \hi\ velocity fields are within $\sim 4.6$~arcsec ($\sim 91$~pc) of the photometric centre.  The best-fitting position angles determined by E2010A varied from $\sim 315$\deg\ to $\sim 285$\deg\ in a roughly linear fashion when moving from the inner to the outer \hi\ disc.  The best-fitting position angles of all fitted models lie within this range.  

The radial variations of the mean orbital speed of the gas as derived using {\sc velfit} ($V_t$, filled and open red circles in Fig.~\ref{velfit_profiles}) agree well with the corresponding circular velocity profiles from E2010A (solid black curves in Fig.~\ref{velfit_profiles}).    All of the final {\sc velfit} models contain significant non-circular velocity components.  The magnitudes of these components are similar for both the bi-symmetric and axisymmetric models, exhibiting similar slight radial variations.  The axisymmetric models suggest the presence of radial flows of $\sim 10-15$~\kms\ for the region $0\lesssim~R\lesssim~100$~arcsec.  Beyond $R\sim 100$~arcsec, the inferred radial velocities increase in magnitude, reaching typical values of $\sim 15$~\kms\ for 120~arcsec~$<R<$~250~arcsec.  The outermost disc ($R>250$~arcsec) contains radial non-circular motions with magnitudes of a few \kms.  The bi-symmetric models fitted to both the high- and low-resolution \hi\ velocity fields suggest the presence of bi-symmetric flows about an axis inclined $\sim 18^{\circ}-30^{\circ}$ relative to the kinematic major axis.  The magnitudes of the $V_{2,t}$ and $V_{2,r}$ components are similar to one another at all radii, increasing from $\sim 8$~\kms\ at $R\lesssim 100$~arcsec to $\sim 15-20$~\kms\ for 100~arcsec~$<R<$~230~arcsec.

\subsubsection{Discussion}
For discussion purposes, we consider only the smoothly-varying radial profiles of the non-circular velocity components fitted to the low-resolution \hi\ velocity field.  These profiles closely match the corresponding profiles of the high-resolution \hi\ velocity field.  Regarding the modeling results, an important question to answer is the following: given that the $V_{2,t}$ $V_{2,r}$ and $V_r$ profiles vary with galactocentric radius in an approximately similar manner, should the optimal radial flow model be preferred over the optimal bi-symmetric model, or vice versa?  

Using {\sc velfit}, \citet{spekkens_sellwood_2007} found $V_r$ of the order of 10~\kms\ throughout the inner $R\lesssim 1.5$~kpc ($\sim 85$~arcsec) portion of the \hi\ disc of NGC~2976.  They argue that such large radial flows will lead to all of the detected gas in the quiescent system being displaced on kpc length-scales on time-scales of 1-3~Gyr.  They therefore disfavour the results of their optimal radial flow model.  While it is true that significant radial flows within a gaseous disc will redistribute the gas over extended time periods, it can be argued that such radial flows are expected near the star-forming cores of galaxies.  Indeed, several authors have linked the effects of stellar feedback to radially expanding gas components in nearby galaxies \citep[e.g.][]{van_eymeren,young_et_al_2003,IC2574}.  For the case of NGC~2915, we therefore do not exclude the possibility of radial flows within the \hi\ disc, especially at inner radii where the effects of star formation may dominate the gas kinematics.  In Elson et al., 2010 (in prep.) we produce detailed three-dimensional models of the \hi\ observations of NGC~2915 in order to link the central \hi\ kinematics to the observed star formation activity.

The \hi\ total intensity map of NGC~2915 (Fig.~\ref{mom0}) clearly displays a bar-like central \hi\ feature, thereby providing strong evidence for an $m=2$ perturbation to the potential of the galaxy.  The stellar disc of NGC~2915 coincides in position, size and orientation to the observed bar-like \hi\ feature.  It can therefore be argued that the $V_{2,t}$ and $V_{2,r}$ radial profiles of the bi-symmetric flow models are reasonable characterisations of \hi\ streaming motions of 0-20~\kms, driven by a bar in the mass distribution.  That the $V_{2,t}$ and $V_{2,r}$ amplitudes are relatively small for $R\lesssim 100$~arcsec is not surprising given the few resolution elements across the bar in this region.  A slightly more subtle, yet arguably more pertinent question to answer, is whether having fixed the position angle of the bi-symmetric distortion to $\phi_b'=322$\deg, i.e. to that of the bar-like feature seen in the \hi\ distribution, leaves the bi-symmetric models insensitive to bi-symmetric distortions that are not aligned with this feature.  To address this concern, additional bi-symmetric flow models were fitted to the low-resolution \hi\ velocity field, this time with $\phi_b'$ allowed to freely vary.   The models were fitted separately to the entire \hi\ disc as well as to only the outer \hi\ disc ($R>200$~arcsec).  Modelling the entire disc yielded $V_{2,t}$ and $V_{2,r}$ radial profiles very similar to those presented in Fig.~\ref{velfit_profiles}c, with the corresponding $V_{2,t}$ and $V_{2,r}$ profiles differing respectively by 3.4~\kms\ and 3.6~\kms\ on average for the cases in which $\phi_b'$ was fixed and allowed to freely vary.  The best-fitting position angle in the plane of the sky for the bi-symmetric distortion is $\phi_b'=319.2$\deg.  The model fitted only to the outer disc has a best-fitting position angle of $\phi_b'=311.9$\deg.  These fitted position angles, being very similar to the position angle $\phi_b'=322$\deg of the observed bar-like feature in the mass distribution, suggest that this feature is indeed driving the elliptical streaming motions of the gas, and that our fitted models are not insensitive to bi-symmetric distortions that are misaligned with it.

\citet{spekkens_sellwood_2007}, using {\sc velfit}, found evidence for bi-symmetric flows of similar magnitude ($\sim 10-20$~\kms) for the case of NGC~2976.  \citet{masset_bureau_2003} used hydrodynamical simulations to show that NGC~2915's observed \hi\ spiral structure can be accounted for by either an unseen bar or a rotating tri-axial dark matter halo.  A problem with the bar scenario is, however, that the mass of the required bar ($\sim 5\times 10^9$~\msun) is very large relative to the total stellar mass of the galaxy ($\sim 8.0\times 10^8$~\msun), thereby making the nature of such a bar problematic.  

\subsection{Method 2: {\sc reswri}}
Our second independent attempt at quantifying the non-circular gas motions in the \hi\ disc of NGC~2915 involves an harmonic decomposition of the line-of-sight-velocities and then, using the theoretical framework laid down by \citet{franx_et_al_1994} and \citet{schoenmakers_1997}, connecting the measured kinematics to perturbations of the potential.  These authors assumed a small stationary distortion of the potential and then used epicycle theory to analytically determine the equations of motion of a test particle.  The equations of motion were used to determine the resulting line-of-sight velocities.  The method of \citet{franx_et_al_1994} and \citet{schoenmakers_1997} is therefore expected to yield reliable results only for small departures from circular motion.  In this sense, {\sc reswri} is not as widely applicable as the {\sc velfit} method which can fit arbitrarily large non-circular velocity components.  It is, however, able to accommodate warped discs with non-constant major axis position angles.  {\sc reswri} does not assume a particular type of potential distortion, unlike the {\sc velfit} technique.

\subsubsection{Fitted models}
Following \citet{schoenmakers_1997}, the line-of-sight velocities are expressed as 
\begin{equation}
V_{los}=V_{sys}+\sum_{m'=1}^N c_{m'}\cos m'\theta + s_{m'}\sin m'\theta,
\label{fourier}
\end{equation}
where $N$ is the order of the fit, $c_{m'}$ and $s_{m'}$ represent the magnitudes of the non-circular line-of-sight velocity components, and $\theta$ is the azimuthal angle in the plane of the galaxy.  Very importantly, because Eqn.~\ref{fourier} is a Fourier series for the line-of-sight velocities, $m'$ represents the harmonic order of the potential perturbation in the \emph{plane of the sky}.  This is unlike $m$ in Eqn.~\ref{fourier_vel} which represents the potential perturbation in the \emph{plane of the disc}.  The two harmonic numbers are related by $m'=m\pm1$.   Following \citet{trachternach_THINGS}, the line-of-sight velocities were expanded up to order $N=3$.

The {\sc gipsy} task {\sc reswri} was used to fit Eqn.~\ref{fourier} to concentric rings in the high-resolution \hi\ velocity field.  Unlike {\sc velfit}, {\sc reswri} does not use the entire observed \hi\ velocity field in a single fit, but rather divides the disc into concentric rings and then separately determines the non-circular line-of-sight velocities for each of them.  As part of the decomposition process, {\sc reswri} first fits a circular tilted ring model to the velocity field to obtain a measure of the orientation parameters of the rings as well as the kinematic centre for each ring.  {\sc reswri} then subtracts the circular rotation model from the observed velocity field and determines the non-circular components of the residual field.  Both \citet{franx_et_al_1994} and \citet{schoenmakers_1997} caution that the centre position of the tilted ring model should always be kept fixed when running {\sc reswri}.  If allowed to vary, the centre will drift so as to make real $c_2$ and $s_2$ harmonics disappear.  We therefore fixed the centres of the rings to the position of the photometric centre mentioned in Sec.~\ref{velfit_results}.  The tilted ring model orientation parameters were allowed to vary freely, since incorrectly fixing them would result in incorrect measures of the harmonic components.  The default uniform weighting was used during the fitting procedure to ensure that points along the minor axis contributed equally to the measured harmonics.  These points contain the most information about non-circular motions.  Ring widths equal to the full width at half maximum of the synthesised beam were used to ensure that adjacent rings were independent of one another.

Rings with radius $R<200$~arcsec were not included in the harmonic decomposition.  The complex gas dynamics within this region are most likely dictated by the ongoing high-mass star formation at the centre of NGC~2915, and therefore cannot be meaningfully interpreted in the context of perturbations of an axisymmetric potential.  Furthermore, the magnitudes of these non-circular motions are expected to be significant fractions of the corresponding circular velocities.  Finally, \citet{schoenmakers_1997} caution that errors in the tilted ring fits are expected to be large in regions where $V(R)\propto R$, thereby preventing a reliable measure of the harmonic components of the velocity field.  For these reasons, only the $R>200$~arcsec portion of the high-resolution \hi\ velocity field was used in the harmonic decomposition.  The portion of the \hi\ disc that we analyse using {\sc reswri} is delimited by the dashed ellipses shown in Fig.~\ref{mom0}.

\subsubsection{Results}\label{reswri_results_section}
\begin{figure*}
	\begin{centering}
	\includegraphics[angle=0,width=2\columnwidth]{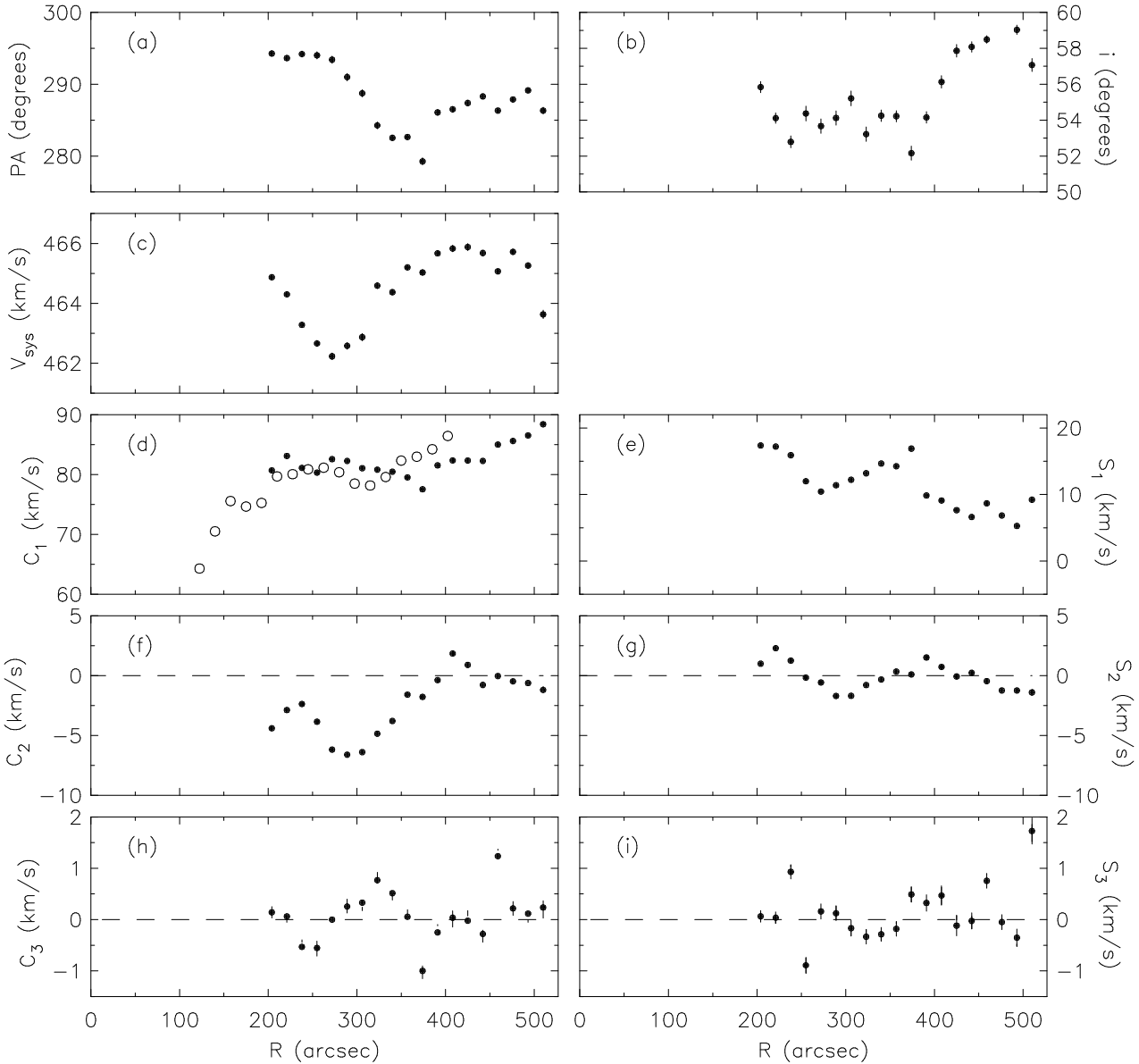}
	\caption{{\sc reswri} harmonic decomposition results (filled circles).  Rows 1 and 2 show the circular tilted ring model parameters fitted by {\sc reswri} to the velocity field.  Rows 3 - 5 show the amplitudes of the harmonic components fitted to the residual velocity.  Error bars represent the formal least-squares errors.  In each panel of rows 4 and 5, the dashed line marks 0~\kms.  The harmonic decomposition was only carried out for radii $R\ge 200$~arcsec.  For comparison, the corresponding $V_t$ profile of the optimal radial flow model fitted by {\sc velfit} to the low-resolution \hi\ velocity field is shown in panel (d) as open circles.}
	\label{reswri_results}
	\end{centering}
\end{figure*}

Figure \ref{reswri_results} shows the results of the harmonic decomposition.  The top three panels show the position angles, inclinations and systemic velocities from the circular tilted ring model fitted to the velocity field by {\sc reswri}.  These fitted parameters agree well with those of the tilted ring model fitted to the velocity field by E2010a.  They are also consistent with the corresponding optimal parameters determined by {\sc velfit} (as listed in Table~\ref{parameters_table}).  {\sc reswri} creates a second residual field by subtracting the fitted harmonic components of each ring from the observed velocity field.  This residual field therefore provides a measure of the signal that was not captured by the harmonic expansion.  For all galactocentric radii larger than 200~arcsec we find a mean absolute residual of 3.6~\kms, suggesting that an harmonic expansion up to third order is capable of capturing most of the non-circular motions.  \citet{trachternach_THINGS} find an average value of 2.9~\kms\ for the median absolute residuals of the 19 THINGS galaxies in their sample.  The average amplitude of the non-circular velocity components for  their entire sample is $6.7\pm 5.9$~\kms.  The corresponding average for NGC~2915 is $11.7\pm4.0$~\kms.

Within the theoretical framework of \citet{franx_et_al_1994} and \citet{schoenmakers_1997}, $c_0$ and $c_1$ correspond respectively to the systemic and circular velocities of each ring, while the higher order terms measure the non-circular velocity components.  A cursory examination of the harmonic decomposition results shows that the $c_3$ term is approximately zero over the full radial range indicating, according to \citet{schoenmakers_1997}, that the correct inclinations were fitted by {\sc reswri}.  \citet{schoenmakers_1997} do caution, however, that in the case of an $m=2$ spiral arm, this inclination is not necessarily equal to the true inclination.  However, in the case of NGC~2915, all of our three independent disc inclination estimates, including the {\sc reswri} estimate, are consistent with one another.  Also shown in Fig.~\ref{reswri_results}d is the corresponding $V_t$ profile of the optimal radial flow model fitted by {\sc velfit} to the low-resolution \hi\ velocity field.

\subsubsection{Discussion}

Perhaps the most striking result from the harmonic decomposition is the large amount of power seen in the $s_1$ component ($s_1\gtrsim$ 10~\kms\ for $R\lesssim400$~arcsec, and then 5~\kms~$\lesssim s_1\lesssim$~10~\kms\  for $R\gtrsim400$~arcsec).  A strong $s_1$ harmonic in the velocity field suggests the presence of an intrinsic $m=2$ velocity distortion, often associated with a tri-axial halo.  The problem, however, is that $m=2$ spiral structure, which is clearly present in the \hi\ distribution, will also result in a significant $s_1$ term in the observed velocity field.  \citet{wong}, using the linearised equations of \citet{canzian_1997} for the velocity perturbations caused by a two-armed spiral density wave, point out that a characteristic kinematic signature in the velocity field due to such a perturbation is a sinusoidal variation with radius of the $s_1$ and $s_3$ harmonic components.  While it may be argued that some sinusoidal variation is present in the $s_1$ term, the $s_3$ term is close to zero for all radii, thereby providing evidence against the perturbational effects of a two-armed spiral density wave to the observed velocity field.  Furthermore, the fact that the bi-symmetric models fitted by {\sc velfit} to the velocity field are insensitive to spiral structure, yet still suggest the presence of sizeable non-circular $V_{2,t}$ and $V_{2,r}$ velocity components, implies that the strong $s_1$ term found by {\sc reswri} cannot solely be due to spiral structure.

The strong $s_1$ term could also be linked to an axisymmetric radial flow.  In their description of the kinematic signatures of different types of non-circular motions, \citet{wong} state that a significant $s_1$ term together with a negligible $s_3$ term can be caused by radial gas inflow.  We indeed see this behaviour in the harmonic terms of NGC~2915.  \citet{wong} go on to list $|ds_3/ds_1|\lesssim 0.1$ over a significant radial range as a further kinematic signature of radial flow.  We measure an average absolute gradient of $|ds_3/ds_1|\approx 5.8\times 10^{-4}$ for the outer disc of NGC~2915, consistent with the radial flow scenario.  A further kinematic signature of radial flow is non-orthogonality of the kinematic major and minor axes.  From the circular tilted ring model fitted by {\sc reswri} to the \hi\ velocity field, the position angle of the kinematic major axis is estimated to be $\sim$~287\deg.  The iso-velocity contours in the \hi\ velocity field suggest the position angle of the kinematic minor axis to be $\sim$~212\deg.  The two kinematic axes thus appear to be non-orthogonal, differing in position angle by only $\sim$~75\deg, thereby further supporting the possibility of radial flow.  This evidence for radial flows is further supported by the fact that {\sc velfit} finds an axisymmetric radial flow model to better match the observed \hi\ velocity field of NGC~2915 than does a bi-symmetric flow model.  This is true regardless of whether the bar angle in the bi-symmetric flow model is kept fixed or is allowed to freely vary.

 \citet{wong} mention that if the $s_1$ term is interpreted as evidence for radial flow, then in a counter-clockwise rotating galaxy, a positive $s_1$ term implies outflow.  Assuming that a galaxy's spiral arms are always trailing, NGC 2915 is rotating counter-clockwise, thereby implying the strong $s_1$ term to be indicative of a radial outflow of the order of $5-17$~\kms.  This result is similar to that of \citet{gentile_ngc3741} who inferred the presence of radial inflows of the order of $5-13$~\kms\ in the \hi\ disc of the nearby dwarf irregular galaxy NGC~3741.  \citet{sancisi_review} hypothesise the existence of NGC~2915's extended \hi\ disc to be as a result of the accumulation of in-fallen pristine gas from the surrounding inter-galactic medium.  This gas, they say, forms ``a reservoir of fresh gas for fuelling star formation in the inner regions''.  The physical process of gas moving from the outer to the inner disc would, however, be associated with an axisymmetric radial \emph{inflow}, and is hence difficult to reconcile with our kinematic evidence for a radial outflow.  \citet{werk_2010} have recently reported on lower- and higher-than-expected oxygen abundances for the respective inner and outer portions of NGC~2915's \hi\ disc.  The authors showed that the few isolated regions of low-level star formation within the outer disc cannot account for the measured oxygen abundance.  They propose a metal-mixing scenario in which metals produced by the star forming core are radially redistributed towards the outer disc.  Such a scenario is consistent with our kinematic evidence for radial outflows.  Furthermore, several authors \citep[e.g.,][]{bureau_1999,masset_bureau_2003} strongly advocate the presence of a bar within NGC~2915's co-rotation radius.  If mass is able to flow along this bar towards the centre of the galaxy, conservation of angular momentum dictates that some mass must also be sent outwards, away from the centre of the system.  Our inferred outward radial motions are consistent with this physical scenario which could perhaps go some way in explaining the existence of NGC~2915's very extended \hi\ disc.
 


A major flaw in the radial flow argument is that the velocity field of a stationary bar potential resembles axisymmetric inflow or outflow.  This, according to \citet{wong},  is because for a flat rotation curve the $s_3$ and $c_3$ terms are expected to vanish in the case of a stationary perturbation to the potential (demonstrated by \citealt{franx_et_al_1994}).  Bars are indeed thought to be the most effective mechanism for transporting gas to the centre of a galaxy \citep{combes_1999}, and so the kinematics alone cannot distinguish between the stationary bar potential and the radial flow scenarios.  \citet{wong} carried out harmonic decompositions of the CO and \hi\ velocity fields of seven nearby spiral galaxies in order to search for evidence of radial gas flows.  They found no unambiguous evidence for radial flows, concluding that ``the inherent non-axisymmetry of spiral galaxies is the greatest limitation to the direct detection of radial flows''.  Thus, although the $s_1$ term provides very suggestive evidence for gas outflow, it should not be treated as conclusive evidence for such gas kinematics.  Elliptical streaming of the gas in a bar-like potential can still contribute significantly to the observed non-circular motions.  The various perturbations to the potential described in this section are summarised in Table~\ref{harmonics_summary}, together with a list of evidence for and against each perturbation type.

\begin{table}
\begin{center}
\caption{Summary of \hi\ velocity field harmonic decomposition results for NGC 2915.}
\begin{tabular}{ccc}
\hline
\\
Perturber 	& Evidence 	& Evidence \\ 
type		& for			& against\\
\\
\hline
\\
$m=2$ spiral 	& Strong $s_1$	& $s_3$ term does	\\
density		& term		& not vary			\\
wave			&			& sinusoidally.		\\
			&			&				\\
			&			&Well-fitted {\sc velfit}\\
			&			&bi-symmetric flow models\\
			&			&are insensitive to	\\
			&			&spiral structure.	\\
\\
\\
Tri-axial		& Strong $s_1$	& Contributions from	\\
halo			& term		& radial flow, and		\\
			&			& $m=2$ perturbation  \\
			&			& due to \hi\ bar.	\\
\\
\\
Axisymmetric	& Strong $s1$ term			& Indistinguishable from	\\
radial flow		& \emph{and} weak $s_3$ term	& elliptical streaming		\\
			&							& in a stationary bar		\\
			& $|ds_3/ds_1|\approx 0$			& potential.				\\
			&							&					\\
			& Non-orthogonal				&				\\
			& kinematic axes.			&					\\
			&						&					\\
			&Well-fitted {\sc velfit}			&					\\
			&axisymmetric flow			&					\\
			& models.					&					\\
\\
\hline
\end{tabular}
\label{harmonics_summary}
\\
\end{center}
\end{table}
\subsubsection{Elongation of the potential}\label{epotential}
NGC~2915 is known to be extremely dark-matter-dominated, with a total mass to $B$-band light ratio of $\sim140$~\msun/$L_{\odot}$ (E2010a). If the non-circular components measured in the velocity field are treated as being due to a tri-axial dark matter halo, they can be used to estimate the axial ratio of the potential.  \citet{schoenmakers_1997} present an expression for the global elongation of the potential which depends on the $s_1$ and $s_3$ harmonics measured by {\sc reswri}.  In the case of NGC~2915, however, the observable signatures of a possible radial outflow and an $m=2$ spiral density wave perturbation to the potential will couple significantly with those of an elongated potential, the axial ratio for which therefore cannot be confidently quantified using the expression from \citet{schoenmakers_1997}.  

As explained in \citet{spekkens_sellwood_2007} and \citet{sellwood_zanmar_sanchez_2010}, the components returned from {\sc velfit} bi-symmetric flow models are \emph{not} sensitive to these confounding factors.  For a mildly distorted, slowly rotating potential and a flat rotation curve, \citet{sellwood_zanmar_sanchez_2010} estimate the axial ratio of the potential as
\begin{equation}
q_{\phi}=\left({V_t\over V_t+2V_{2,r}}\right)^{1/2}.
\end{equation}
We use the $V_t$ and $V_{2,r}$ profiles from our bi-symmetric flow models fitted to the low-resolution \hi\ velocity field to estimate $q_{\phi}$.  We do this for the model in which $\phi_b'=322$\deg\ was kept fixed, and for the model in which it was allowed to freely vary (attaining a best-fitting value of $\phi_b'=319.2$\deg).  

\begin{figure}
	\begin{centering}
	\includegraphics[angle=0,width=1\columnwidth]{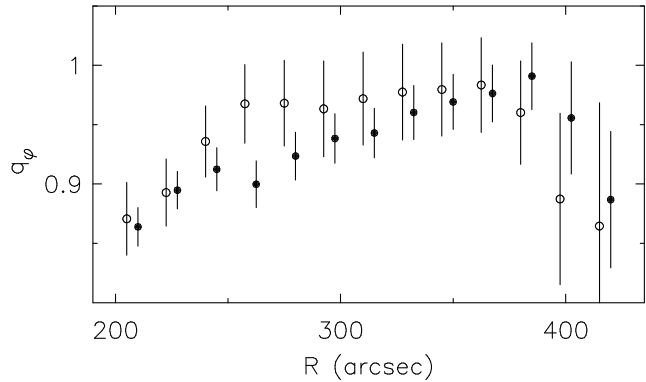}
	\caption{Radial distribution of the axial ratio of the potential of NGC~2915.  Closed and open circles correspond to our fitted bi-symetric flow models in which $\phi_b'$ was held fixed and allowed to freely vary respectively.  For the sake of clarity, the open circles have been shifted by $-5$~arcsec.  Error bars were estimated using Gaussian error propagation of the non-parametric estimates of the true uncertainties in $V_t$ and $V_{2,r}$, generated by applying a bootstrap method.}
	\label{epot}
	\end{centering}
\end{figure}

The two models yield very similar $q_{\phi}$ radial profiles, shown in Fig.~\ref{epot}.  In both cases, the axial ratio for the potential is close to unity for nearly all radii, generally being $\gtrsim 0.9$.  Both bi-symmetric flow models have slowly varying $V_{2,t}$ and $V_{2,r}$ profiles for 250~arcsec~$\lesssim R \lesssim$~400~arcsec.  Additionally, the amplitudes of the non-circular velocity components are less than 10~percent of the corresponding circular velocities.  We therefore treat the $q_{\phi}$ estimates within this radial range as being the most reliable.  Average axial ratios of $q_{\phi}=0.92\pm 0.21$ and $q_{\phi}=0.92\pm 0.35$ are obtained within this radial range for the models in which $\phi_b'$ is fixed and allowed to vary respectively.  These estimates are consistent with an axial ratio of $q_{\phi}=1$, and are consistent with the estimated lower limit of $q_{\phi}\gtrsim 0.98$ from \citet{sellwood_zanmar_sanchez_2010} for the axial ratio of NGC~3198.  We stress, however, that Eqn.~\ref{epot} \emph{will not} yield reliable $q_{\phi}$ estimates when the potential is significantly distorted or when it is fast-rotating.  Neither of these possibilities can be confidently ruled out for the case of NGC~2915.  While our results are strongly suggestive of a perfectly round halo for NGC~2915, other possibilities cannot be excluded due to the limitations of the method discussed above.

\section{Summary and Conclusions}\label{conclusions}
We have searched for non-circular flows within the extended \hi\ disc of NGC~2915 using two independent methods.  The resulting non-circular velocity components have been interpreted in the context of axisymmetric and non-axisymmetric perturbations to the system's gravitational potential.  The results allow us to probe the gravitational potential of the system out to radii far beyond the edge of the stellar disc, into the regions of the galaxy that are completely dark-matter-dominated.  

The first implemented routine, {\sc velfit}, fits a specified kinematic model to the entire \hi\ velocity field.  Models were fitted to both high- and low-resolution versions of the observations, the corresponding results of which are consistent with one another.  Axisymmetric radial flow models best match the observed \hi\ velocity field.  The optimal radial flow models suggest the presence of radial flows as large as 20~\kms\ over a large fraction of the \hi\ disc.  We argue that, at least at inner radii, these inferred radial flows could be the result of the deposition of kinetic energy into the inter-stellar medium by massive stars and supernovae explosions.  {\sc velfit} was also used to fit bi-symmetric flow models to the \hi\ velocity field.  The optimal models suggest bi-symmetric flows of $\sim 8$-20~\kms\ throughout most of the \hi\ disc.  These flows occur along an axis oriented $\sim 18^{\circ}$-$30^{\circ}$ relative to the major axis of the disc, thereby linking them to the central bar-like feature seen in the mass distribution of NGC~2915.  These flows could therefore be caused by streaming gas motions within a bar-like perturbation of the potential.

NGC~2915 is extremely dark-matter-dominated.  The $V_t$ and $V_{2,r}$ profiles from our fitted bi-symmetric flow models have been used to estimate the axial ratio of the potential under the assumption that it is mildly distorted and slow-rotating.  Our results are consistent with the potential being perfectly round, with an average axial ratio of $q_{\phi}\sim 0.92$.  We stress, however, that because of the underlying assumptions used to obtain this result, it cannot be treated as indisputable evidence against a possibly aspherical halo.

The second routine implemented to quantify the non-circular motions is {\sc reswri}.  Evidence is found for an $m~=~2$ term in the \hi\ velocity field.  The possibility of elliptical streaming in a stationary bar-like potential detracts from the otherwise good evidence that is provided by the strong and weak $s_1$ and $s_3$ terms for an axisymmetric radial outflow of the order of 5-17~\kms.  Such outflows, $\sim 6$-20~percent of the outer \hi\ disc rotation velocity, do not provide intuitive insight into the formation history of the \hi\ disc if it is thought of as having accumulated from in-fallen pristine gas from the surrounding inter-galactic medium.  Under such circumstances, the in-fallen gas would be associated with a radial inflow rather than an outflow as it is transported to the central regions of the galaxy.  It is possible, however, that some material is being re-distributed into the outer disc in order to conserve angular momentum as other material flows inwards along a bar.  A clear understanding of the formation history of NGC~2915's extended \hi\ disc is nonetheless still elusive.  The limitation on the search for indirect evidence of radial flows set by the inherent non-axisymmetry of the galaxy gives rise to the need for deeper \hi\ observations of NGC~2915's outer disc in order to directly detect possible cold gas accretion processes and to carefully search its local environment.

\section{Acknowledgements}\label{acknowledgements}
The work of ECE is based upon research generously supported by the South African SKA project.  All authors acknowledge funding recieved from the South African National Research Foundation.  The work of WJGdeB is based upon research supported by the South African Research Chairs Initiative of the Department of Science and Technology and the National Research Foundation.  The Australia Telescope Compact Array is part of the Australia Telescope which is funded by the Commonwealth of Australia for operation as a National Facility managed by CSIRO.  Finally, all authors thank the referee, Kristine Spekkens, for her constructive comments towards the paper.

\label{lastpage}

\end{document}